# COCCIALAB

*To discover the causes of social, economic and technological change*

# Cyclical phenomena in technological change

Mario COCCIA
CNR -- National Research Council of Italy

# Cyclical phenomena in technological change

*Mario Coccia[1]*

CNR -- National Research Council of Italy

*Contact: E*-mail: mario.coccia@cnr.it

Mario Coccia 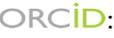: http://orcid.org/0000-0003-1957-6731

## Abstract

The process of technological change can be regarded as a non-deterministic system governed by factors of a cumulative nature that generate cyclical phenomena. In this context, the process of growth and decline of technology can be systematically analyzed to design best practices for technology management of firms and innovation policy of nations. In this perspective, this study focuses on the evolution of technologies in the U.S. recorded music industry. Empirical findings reveal that technological change in the sector under study here has recurring fluctuations of technological innovations. In particular, cycle of technology has up wave phase longer than down wave phase in the process of evolution in markets before it is substituted by a new technology. Results suggest that radical innovation is one of the main sources of cyclical phenomena for industrial and corporate change, and as a consequence, economic and social change.

---

[1] The author declares that he has no relevant or material financial interests that relate to the research discussed in this paper.

# INTRODUCTION

The process of technological change is not random in its evolutionary pathways and the existence of recurring fluctuations in the evolution of innovations is not unexpected (Sahal, 1981). A theoretical framework for to analyze systematically the growth and decline of innovations is critical for technological analysis, management of technology and subsequent competitive advantage of firms and nations (Coccia, 2017, 2019d, 2020c). In this perspective, this paper has *two goals*. The *first* is to measure the relative growth of disruptive technology versus established technologies in markets. The *second* goal is to analyze the technological cycle of disruptive innovations and suggest properties that can explain and generalize the behavior of disruptive technologies and technological change in markets.

This study is part of a large body of research on the evolution of technology to explain, with a new perspective, technological and economic change in society (Coccia, 2005, 2005a, 2005b, 2006, 2017, 2019, 2019a, 2019b, 2020c; Coccia and Watts, 2020). This study here focuses on the evolution of disruptive technologies[2] that sustain industrial and corporate change (Coccia, 2017a, 2017c, 2018b, 2019d). Abernathy and Clark (1985, pp. 12-13, original emphasis) claim that a type of innovation: "disrupts and renders established technical and production competence obsolete …. It thus seems clear that the power of an innovation to unleash Schumpeter's 'creative destruction' must be gauged by the extent to which it alters the parameters of competition, as well as by the shifts it causes in required technical competence". Christensen and Bower (1996) argue that disruptive technologies have a new set of product/process features from those associated with mainstream technologies. In the initial stage of development, disruptive technologies serve niche segments that value their new kinds of attributes. Subsequently, the evolution of disruptive technologies increases the technical performance to satisfy mainstream customers (cf., Vecchiato, 2017). The study by Christensen (1997) identifies specific characteristics: *a)* the performance trajectory provided by disruptive technology; *b)* the performance trajectory demanded by mainstream market. Christensen et al. (2015) claim that disruptive innovations are generated by small firms with fewer resources that successfully challenge established incumbent businesses. In particular, new entrants endeavor to develop disruptive technologies in market segments, delivering the market performance that incumbents' mainstream customers require (Christensen et al., 2015; Christensen, 1997; Foster, 1986). Innovative firms, generating disruptive technologies, grow more rapidly than other ones (Abernathy and Clark, 1985; Tushman and Anderson, 1986, p. 439). Christensen (1997) also states that disruptive innovations generate significant shifts in markets (cf., Henderson, 2006; Coccia, 2019e). These market shifts are due to competence-destroying and competence-enhancing because some firms can either destroy or enhance the competence existing in industries (cf., Coccia, 2018; Hill and Rothaermel, 2003; Tushman and Anderson, 1986). In general, new firms generate competence-destroying discontinuities that increase the environmental turbulence with disruptive technologies, whereas incumbents focus mainly on competence-enhancing discontinuities, improving their products and services in markets (usually most profitable), that decrease this turbulence in markets. Moreover, disruptive innovations change habits of consumers in markets and undermine the competences and complementary assets of existing producers (Markides, 2006, pp.22-23; Zach et al., 2020). Some scholars argue that the feature of disruptiveness of innovations is distinct from the radicalness or the competency-destroying dimensions of innovations (Govindarajan and Kopalle, 2006).

---

[2] This study uses the concepts of disruptive technologies and disruptive innovations interchangeably.

Overall, then, disruptive technologies generate a disruptive creation and technological change in markets and society (Coccia, 2020). Although several contributions in these fields of research, the patterns of disruptive innovations that generate structural change in markets are hardly known. The main aim of this article is to explain and generalize, whenever possible, the behavior and characteristics of disruptive technologies within industrial competition. In particular, this study addresses some basic questions for economics of innovation:

- how to measure the *growth rate* of disruptive technologies that gives a precise evaluation and prediction of the speed of substitution versus existing technologies?
- what is the shape of technological cycle in disruptive innovations in the long run?

Next sections confront these questions here by proposing an analytical approach, based on allometric model, which endeavors to explain the evolution of disruptive technologies in competitive markets and suggest some properties of technological change induced by these vital technologies.

# THEORETICAL BACKGROUND

Technological change generates the emergence of usable innovations that can transform the demand and supply in existing markets, creating new businesses and markets with a significant impact in society (Arthur, 2009; Coccia, 2017; Gilbert, 2003; Hall and Martin, 2005; Hosler, 1994). Technological change is driven by different typologies of innovations (Coccia, 2005; Govindarajan and Kopalle, 2006). One of the most important types is disruptive innovations that have a significant and far-reaching impact in markets and society (Coccia, 2017a, 2018; Danneels, 2004, 2006; Hargadon, 2003; Yu and Hang, 2010). Adner (2002, pp. 668-669) claims that:

> Disruptive technologies . . . introduce a different performance package from mainstream technologies and are inferior to mainstream technologies along the dimensions of performance that are most important to mainstream customers. As such, in their early development they only serve niche segments that value their non-standard performance attributes. Subsequently, further development raises the disruptive technology's performance on the focal mainstream attributes to a level sufficient to satisfy mainstream customers.

Guo et al. (2019) argue the multidimensional nature of disruptive innovations (cf., Christensen and Bower, 1996; Christensen and Overdorf, 2000; Nagy et al., 2016; Schmidt and Druehl, 2008). In particular, disruptive technologies can generate: 1) localized disruption within a market or industry; 2) disruption with larger influences, affecting many industries and changing habits of customers, societal norms and institutions (cf., Schuelke-Leech, 2018, p. 261; Coccia, 2019c; Van de Ven and Garud, 1994). In fact, Schuelke-Leech (2018, p. 261) argues that disruptive technologies can have a minor impact localized within a market or industry or a major effect on many industries and institutions generating socioeconomic change. Kikki et al. (2018) claim that a disruptive creation is an event in which an agent must redesign its strategy to survive a change in turbulent markets (cf.,

Li et al., 2018; Mahto et al., 2017; Majumdar et al., 2018). In this context, Calvano (2006) argues that: "we highlight the role of destruction rather than creation in driving innovative activity. The formal analysis shows that destructive creation unambiguously leads to higher profits whatever the innovation cost" (cf., Tripsas, 1997). Disruptive technologies disturb the business models of incumbents that have to counter mobilize resources to sustain their competitive advantage in the presence of market change (Garud et al., 2015). Kessler and Chakrabarti (1996, p. 1143) argue that disruptive: "innovation speed (a) is most appropriate in environments characterized by competitive intensity, technological and market dynamism, and low regulatory restrictiveness; (b) can be positively or negatively affected by strategic-orientation factors and organizational-capability factors; and (c) has an influence on development costs, product quality, and ultimately project success". Zach et al. (2020) show that adoption speeds, given by first vs. late adoption, make a difference as the former are awarded a significant increase in market value. Love et al. (2020) argue that the rate of digital disruption is placing increasing pressure on organizations to adopt emerging technologies that improve their productivity and bottom-lines. These characteristics of disruptive technologies require that incumbents undertake specific R&D investments (Christensen and Raynor, 2003; Radnejad and Harrie, 2019). In this context, Vecchiato (2017) explores why incumbent firms fail to identify new markets in the presence of disruptive technologies and why incumbents can lose their leadership: a reason may be the inability to recognize either the rising 'social' market, where customers use products for fulfilling their need for friendship, or the 'esteem' market, where customers use products for fulfilling their need for achievement. Park (2018) has extended the theory of disruptive innovation by analyzing the behaviors of incumbents and new entrants to establish a competitive technology strategy in markets. Reinhardt and Gurtner (2018) argue that, in disruptive innovation theory, performance dimensions and price influence adoption depend on product category (cf., Coccia, 2016). However, these scholars do not find systematic differences between the effects of more technologically-oriented performance dimensions or price on adoption. Moreover, product embeddedness, defined as the degree to which a product is anchored in the social, market and technological system of user, is an important moderator that complements extant theory and explains the dynamics of disruptive innovations.

The behavior of disruptive technologies in markets is associated with their evolution that can be explained with theories based on processes of competitive substitution of a new technology for the old one (Fisher and Pry, 1971; Sahal, 1981; Utterback et al., 2019). Coccia (2019) argues that emerging technologies often supplant for more mature technologies in markets. This dynamic behavior between technologies leads to the dominance of a new technology on another one in markets (cf., Berg et al., 2019). A model that operationalizes the competition between technologies is by Fisher and Pry (1971). This model proposes that a new product/process can be a substitute for a prior one, such as synthetic *vs.* natural fibers, synthetic *vs.* natural rubber, etc. (cf., Fisher and Pry, 1971, p. 77; Sahal, 1981). The history of technology has many examples of competition between disruptive technologies and established ones, such as the diffusion and dominance of steamship as efficient means of transportation of goods and people vs. sailing ship (Sahal, 1981, p. 79ff). Another example of disruptive technology is the diffusion of Solvay process that in the 1900s destroys the Leblanc process in the manufacturing sector of soda (Freeman, 1974). Coccia (2018) describes some examples from the market of devices for data storage in which Universal Serial Bus (USB) technology has destroyed the use of 3.5-inch floppy disks, Compact Disc, etc. generating industrial and corporate change (cf., Coccia, 2015, 2017a,2017b, 2017c, 2018, 2019). Bock (2015) describes the technological disruption in construction automation, which is driven by upcoming ubiquity of robotics. Zhang at al. (2019) analyze the approach of disruptive innovations by the Haier group in the industry of home appliances and consumer electronics. Finally, Coccia (2020) shows disruptive technologies in cancer imaging given by deep learning methods that can generate a shift of technological paradigm for diagnostic assessment of any cancer type and disease. In general, disruptive technologies have the characteristic of substitutes with a powerful force in markets to improve products and processes that generate technical, economic and social change (Porter, 1980).

## METHODOLOGY AND STUDY DESIGN

- *Evolutionary model for measuring the growth of creative disruption*

The first goal of this study is to measure and analyze the growth rate of disruptive technologies, extending previous results by Coccia and colleagues. This paper here proposes a simple model of growth of a disruptive technology in relation to an established one (cf., Coccia, 2019f). First of all, disruptive technology is a radical innovation, based on new products and/or processes, having high technical performance, which destroy gradually the usage value of all established techniques and/or devices previously sold in society (Coccia, 2017c). The proposed model here is based on the biological principle of allometry that was developed in zoology to study the differential growth rates of the parts of a living organism's body in relation to the whole body (cf., Reeve and Huxley, 1945;

Coccia, 2019f). Sahal (1981) applies this model to explain patterns of spatial diffusion of technology, providing interesting case studies in the agriculture, manufacturing, steel production, electricity generation, etc.

The suggested model is based on following assumptions, described in Coccia (2019f):

(1) Suppose the simplest possible case of only two technologies, *V* (*established technology*) and *Kl* (a *disruptive technology*).

(2) Let *Kl(t)* be the level of a disruptive technology *Kl* at the time *t* and *V(t)* be the level of an established technology *V* at the same time.

Suppose that both *Kl* and *V* evolve according to a *S*-shaped pattern of technological growth, which is given by the differential equation of logistic function. For *V*, *established technology*, the starting equation is:

- $$\frac{1}{V}\frac{dV}{dt} = \frac{b_1}{K_1}(K_1 - V)$$

The equation can be rewritten as:

- $$\frac{K_1}{V}\frac{1}{(K_1 - V)}dV = b_1 dt$$

The integral of this equation is:

- $$\log V - \log(K_1 - V) = A + b_1 t$$

- $$\log \frac{K_1 - V}{V} = a_1 - b_1 t$$

- $$V = \frac{K_1}{1 + \exp(a_1 - b_1 t)}$$

$a_1 = b_1 t$ and $t =$ abscissa of the point of inflection.

The growth of $V(t)$ can be described respectively as:

$$\log \frac{K_1 - V}{V} = a_1 - b_1 t \qquad [1]$$

*Mutatis mutandis*, for disruptive technology $Kl(t)$ the equation is:

$$\log \frac{K_2 - Kl}{Kl} = a_2 - b_2 t \qquad [2]$$

The logistic curve is a symmetrical *S*-shaped curve with a point of inflection at 0.5K, with $a_{1,2} =$ constants depending on initial conditions, $K_{1,2} =$ equilibrium levels of growth, and $b_{1,2} =$ rate-of-growth parameters (1=established technology=$V$; 2=disruptive technology=$Kl$).

Solving equations [1] and [2] for $t$, the result is:

$$t = \frac{a_1}{b_1} - \frac{1}{b_1} \log \frac{K_1 - V}{V} = \frac{a_2}{b_2} - \frac{1}{b_2} \log \frac{K_2 - Kl}{Kl}$$

The expression generated is:

$$\frac{V}{K_1 - V} = C_1 \left( \frac{Kl}{K_2 - Kl} \right)^{\frac{b_1}{b_2}} \qquad [3]$$

In equation [3], $C_1 = exp[b_1(t_2-t_1)]$ with $a_1=b_1 t_1$ and $a_2=b_2 t_2$ (cf. Eqs. [1] and [2]); applying mathematical transformations, the evolutionary growth of disruptive technology in relation to established technology is given by:

$$Kl = A \, (V)^B \qquad [4]$$

where $A = \dfrac{K_2}{(K_1)^{\frac{b_2}{b_1}}} C_1$ and $B = \dfrac{b_2}{b_1}$

The logarithmic form of the equation [4] is a simple linear relationship:

$$\log Kl = \log A + B \, \log V \qquad [5]$$

*B* is the coefficient of growth that measures the evolution of disruptive technology *Kl* in relation to established technology *V*.

This model [5] has linear parameters that are estimated with the Ordinary Least-Squares Method. The value of *B* in model [5] measures the relative growth of *Kl* in relation to the growth of *V*. The coefficient *B* indicates different patterns of technological evolution in markets, namely:

- *B* < 1, whether new technology *Kl* destroys at a *lower* relative rate of change established technology *V* (*low growth of disruptive technology*)

- *B* = 1, then the disruptive technology *Kl* substitutes established technology *V* at a *proportional* rate of change (*proportional growth of disruptive technology*)

- *B* > 1, whether disruptive technology *Kl* destroys established technology V at *greater* relative rate of change over the course of time (*acceleration of disruptive technology in markets*)

- *Data and their sources*

The empirical analysis is based on data of technologies in recorded music industry (e.g., vinyl, 8 track, cassette, CD and streaming technology) in the USA, 1973-2019 period. The U.S. national system of innovation is a vital case study to show general patterns of the evolution of new technologies in advanced economies (Steil et al., 2002). Source of data for recorded music technology is Recording Industry Association of America (RIAA), which provides data on U.S. recorded music revenues and music sales volumes from 1973-2019 (RIAA, 2019, 2020). Note that the first year represented in dataset is not the year of invention of technologies under study (cf., RIAA, 2019).

- *Measures*

Theory construction in strategic management must encompass reliable and valid measures for key innovation characteristics (Govindarajan and Kopalle, 2006). This study applies the following measures to detect the creative disruption of new technologies.

◻ *Competition between CD and cassette technology in U.S. recorded music industry*

Measures are:

− Recorded music revenues in millions $ (adjusted for inflation, 2018 Dollars) of CD technology is a measure of the evolution of disruptive-new technology ($Kl$)

− Recorded music revenues in millions $ (adjusted for inflation, 2018 Dollars) of cassette technology is a measure of the evolution of established-old technology ($V$)

◻ *Competition between streaming and CD technology in U.S. recorded music industry*

Measures are:

− Recorded music revenues in millions $ (adjusted for inflation, 2018 Dollars) of streaming technology is a measure of the evolution of this disruptive technology ($Kl$). Note that streaming technology is measured here including different modes: paid subscription, on-demand streaming,

other Ad-supported streaming, sound exchange distributions and limited tier paid subscription.

− Recorded music revenues in millions $ (adjusted for inflation, 2018 Dollars) of CD technology is a measure of the evolution of this established technology ($V$) in the period under study.

*Remark*: values are at recommended or estimated list price (cf., RIAA, 2019).

- *Data analysis procedure for investigating the growth of disruptive technologies and the shape of technological cycle*

The operationalization of the proposed model [5] in the case study here is specified as follows:

$$Log\ K_{l_t} = log a + B\ log\ V_t + u_t \qquad [6]$$

$a$ is a constant; *log* has base $e = 2.7182818$; $t$ = time; $u_t$ = error term

$K_{l_t}$ is a measure of the growth of disruptive technology in U.S. recorded music industry

$V_t$ is a measure of the growth of established technology in U.S. recorded music industry

The relationship [6] is analyzed using ordinary least squares (OLS) method for estimating the unknown parameters in a linear regression model. Statistical analyses are performed with the Statistics Software SPSS® version 24.

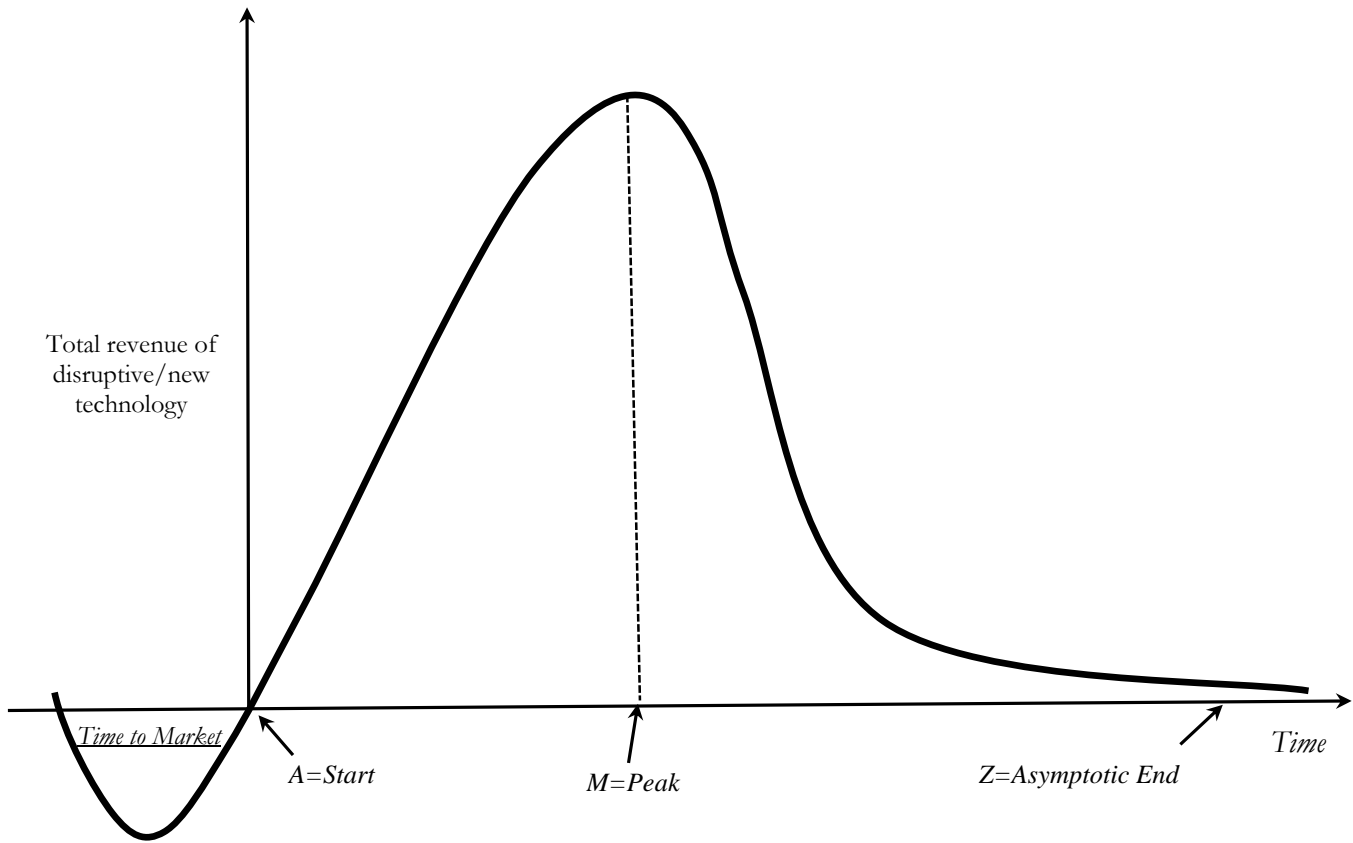

*TECHNOLOGICAL CYCLE*

This study also proposes some equations for technology analysis of the technological cycle of disruptive innovations.

Let
*i*  = technology *i*
A   = year of the starting of revenues of technology *i* in market
M   = year of the peak of revenues of technology *i* in market
Z   = year of ending of revenues of technology *i* in market

The technology analysis of the cycle of disruptive innovation is given by following equations:

- $AZ_i$ = length of the cycle of technology *i*
  $$AZ_i = Z_i - A_i \qquad [7]$$

- $AM_i$ = length of the up wave phase of technological cycle $i$
  $$AM_i = M_i - A_i \qquad [8]$$
- $ZM_i$ = length of the down wave phase of technological cycle $i$ (also called disruption period $DP_i$ of technology $i$)
  $$ZM_i = Z_i - M_i \qquad [9]$$
  $$ZM_i = DP_i = \text{Disruption period of technology } i = Z_i - M_i$$

## RESULTS

*1.1 A short history of technology in recorded music industry*

Recorded music industry in the United States of America is an interesting case study. From 1973 to 2019, the technological trajectories for delivering sound−included music−have had radical changes (cf., RIAA, 2019, 2020). The earliest disc records (1880s–1890s) were made of variety of materials including hard rubber. In 1930, the Radio Corporation of America (RCA) by the Victor Talking Machine Company launched the first commercially-available vinyl long-playing record. Vinyl had lower playback noise level than shellac (Mudge and Hoek, 2001). During and after World War II when shellac supplies were extremely limited, some 78 RPM (Revolutions per Minute) records were pressed in vinyl for distribution to US troops. Since 1939, Columbia Records continued the development of this technology and in 1948 released their 33 ⅓ RPM, which is made from polyvinyl chloride and pressed on a 12" diameter flexible plastic disc. The sound is recorded in the grooves in the vinyl and while the record spins, the needle runs along the grooves and passes the information to the electromagnetic head (Read, 1952). After vinyl, the 8-track tape (formally Stereo 8) was a magnetic tape sound-recording technology popular in the United States from the mid-1960. The Stereo 8 Cartridge was created in 1964 by William Lear, using previous technology of tape cartridge introduced in 1958 by the Radio Corporation of America-RCA-Records Label to be used in cars in a phase of growing production in automotive industry. Subsequently, during the 1970s and 1980s period, the most common technological device to deliver music was compact cassettes based on analog magnetic tape for audio recording and playback. This product innovation was developed by Philips company, released in 1962 and introduced in the U.S. market in 1964. Ray Dolby develops in 1968 a technology called Dolby noise reduction to increase the sound quality of cassette tapes. These technological advances led cassette tapes to be a dominant technology on 8-track tapes in the mid-1970s and in the early 1980s. However, the emerging technology of compact audio disc (CD), co-developed by Philips and Sony corporations and launched in 1982, generates a further market shift (BBC News, 2007). CD technology is a digital optical disc originally developed to store and play only sound recordings but it was later also adapted for storage of other data (Coccia, 2018). In the mid-1990s and in the early 2000s, the high sound quality of CD led this technology to be the dominant one in markets, overtaking cassette sales from 1991 to 2005 period (cf., RIAA, 2019). The revolution of Information and Communication Technologies (ICTs) has generated other new technologies for market of recorded music, based on transmission of video/audio information over the Internet, such as:

*Download mode.* The content file is completely downloaded and then played. This mode requires long downloading time for the whole content file and needs a large hard disk space (a consequential problem that supports technical advances towards streaming mode, cf., Coccia, 2017).

*Streaming mode.* The content file is not required to be downloaded completely and it plays while parts of the content are being received and decoded. Files seen can be inserted in favorite items and video-sharing platform remembers your preferences for future necessities and desires. In particular, the video streaming technology delivers audio and video over the Internet network to reach many customers using their personal computers, personal digital assistants, smartphones or other Information and Communication Technology (ICT) devices. The growth of streaming technology is due to broadband networks, efficient techniques of video and audio compression, a higher quality and variety of audio and video services over Internet, etc. A streaming media player can be either an integral part of a browser, a plug-in, a separate program, or a dedicated device, such as Apple TV, iPod, etc. For streaming technology, the UDP/IP (User Datagram Protocol/ Internet Protocol) is used to deliver the multi-media flow as a sequence of small packets. The application of layer protocol RTP/RTSP (Real-time Transport Protocol /Real Time Streaming Protocol), which is implemented on top of UDP/IP, provides an end-to-end network transport for video streaming.

There are different modes of streaming video content distribution, such as (cf., RIAA, 2019):

- *Sound Exchange Distributions* based on payments to performers and copyright holders for digital radio services under statutory licenses
- *Paid Subscription* includes streaming, tethered, and other paid subscription services not operating under statutory licenses
- *Limited Tier Paid Subscription* includes streaming services with interactivity limitations by availability, device restriction, catalog limitations, on demand access, or other factors
- *On-Demand Streaming* includes Ad-supported audio and music video services not operating under statutory licenses
- *Other Ad-supported Streaming* includes revenues paid directly for statutory services that are not distributed by Sound Exchange and not included in other streaming categories.

Coelho (2019) argues that new technology of digital distribution is revolutionizing the pop-rock music market and shows *how* firms respond to this disruptive technological revolution and how this response changes strategic management. Lee et al. (2016) claim that the Internet has changed the media contents industry from records into online digital products. Essling et al. (2017) show how the advent of digitization has changed firm strategy and that record labels release more singles with shorter intervals in between when facing greater competitive pressure. Aguiar (2017) points out that the growth in interactive music streaming is raising questions about its effects on music industry, such as if streaming enhances product discovery, and if consumers value mobility, then free streaming could stimulate the use of channels that allow mobile consumption (cf., Blanc and Huault, 2014; Cavazos and Szyliowicz, 2011). Wlömert and Papies (2016) argue that on-demand streaming services, which rely on subscription fees or advertising as a revenue source (e.g., Spotify), are a topic of ongoing controversial debate in music industry because their addition to the distribution mix entails the risk of cannibalization of other distribution channels (e.g., purchases of downloads or CDs) and might reduce overall revenues. However, results show that the net effect of paid streaming services on revenue is positive. In particular, at the industry level, findings suggest that the negative effect of free streaming on industry revenue is offset by the positive effect of paid streaming (Wlömert and Papies, 2016). Overall, then, many studies reveal that online streaming positively impact music record sales (cf., Steininger and Gatzemeier, 2019).

## 1.2 The growth of disruptive technologies in recorded music industry in the United States of America

− CD as disruptive technology of cassette technology

**Table 1** Parametric estimates of the model of disruptive CD technology on established cassette technology, 1984-2008 period in U.S. market

| | *Constant* $\alpha$ *(St. Err.)* | *Coefficient* $\beta=B$ *(St. Err.)* | *$R^2$ adj.* *(St. Err. of the Estimate)* | *F* *(sign.)* |
|---|---|---|---|---|
| *Dependent variable*: *log* annual recorded music revenues of disruptive CD technology (value adjusted for inflation, 2018 dollars) | | | | |
| CD *vs.* cassette technology | −9.8* (4.72) | 2.1*** (0.55) | 0.51 (0.64) | 14.38 (0.003) |

*Note*: *** significant at 1‰; * significant at 1%; Explanatory variable is *log* annual recorded music revenues of cassette as established technology (value adjusted for inflation, 2018 dollars)

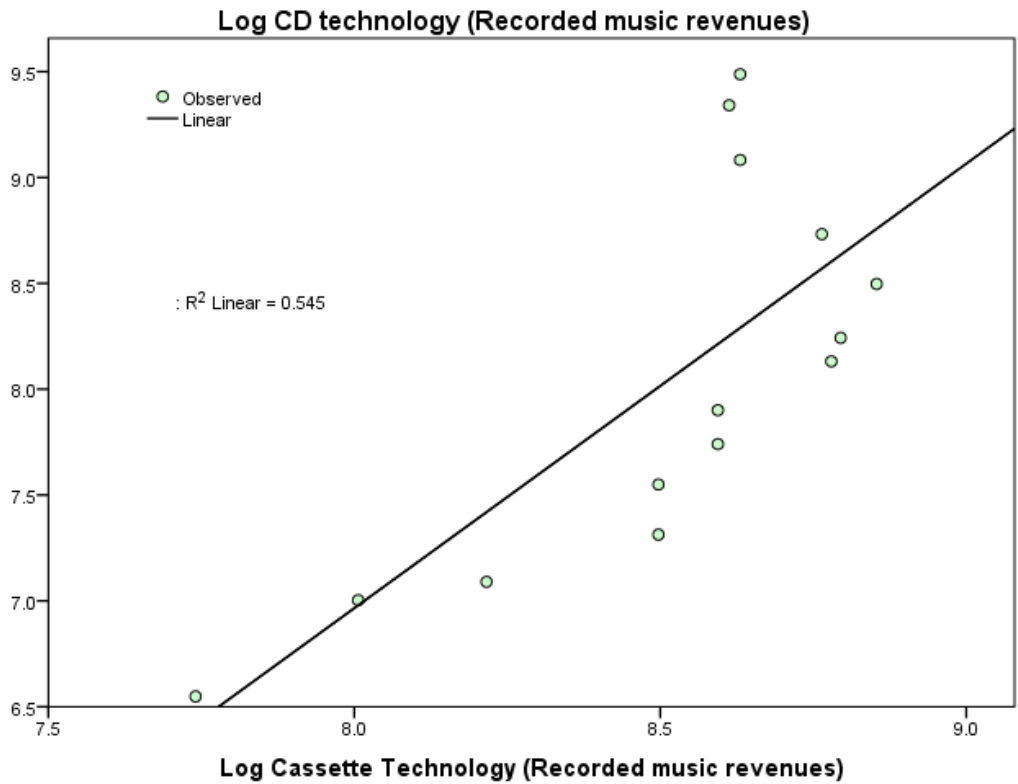

**Figure 1.** Fit line of the growth of recorded music revenues in CD technology (*disruptive technology*) on cassette technology (*established technology*), 1984-2008 period in U.S. market (*log-log* scale; value adjusted for inflation, 2018 dollars)

− Streaming technology as disruptive technology of CD technology

**Table 2** Parametric estimates of the model of disruptive streaming technology on CD technology, 2004-2018 period in U.S. market

| | Constant $\alpha$ (St. Err.) | Coefficient $\beta=B$ (St. Err.) | $R^2$ adj. (St. Err. of the Estimate) | F (sign.) |
|---|---|---|---|---|
| Streaming technology | 17.22*** | −1.28*** | 0.95 | 240.01 |
| vs. CD technology | (0.67) | (0.08) | (0.27) | (0.001) |

*Note*: *** significant at 1‰; Explanatory variable is *log* annual recorded music revenues of CD technology as established technology (value adjusted for inflation, 2018 dollars). Streaming technology is measured here including recorded music revenues of different modes: paid subscription, on-demand streaming, other Ad-supported streaming, sound exchange distributions and limited tier paid subscription.

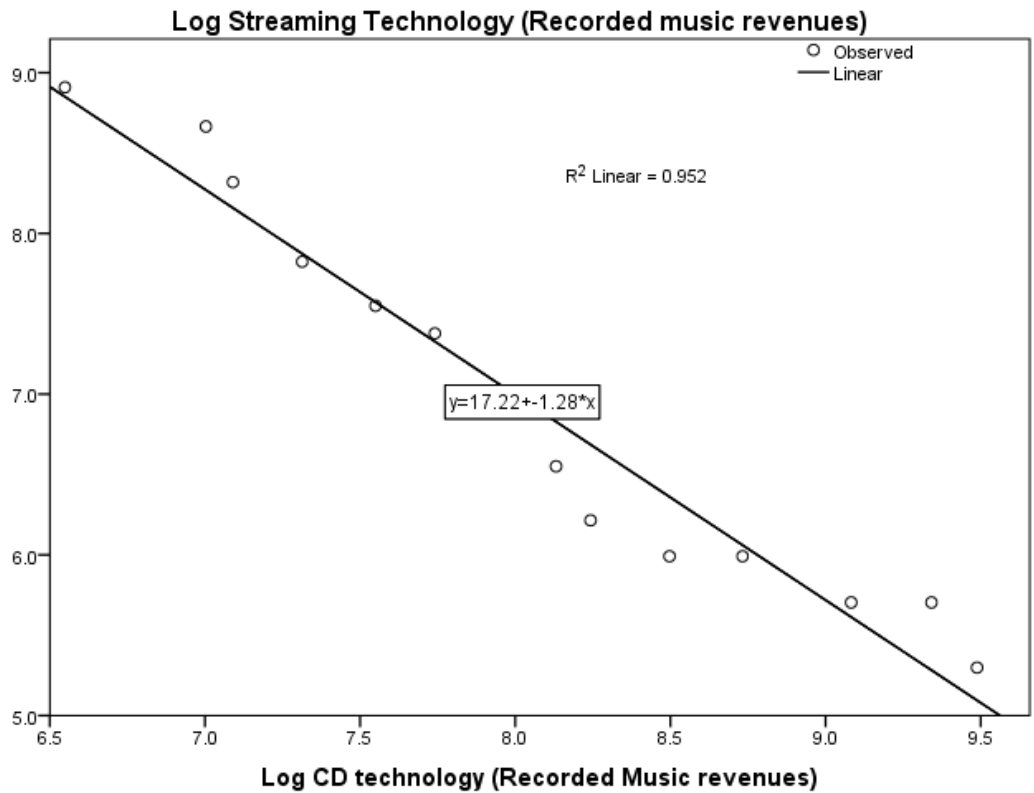

**Figure 2.** Fit line and estimated relationship of the growth of recorded music revenues of disruptive streaming technology on CD technology, 2004-2018 period in U.S. market (*log-log* scale; value adjusted for inflation, 2018 dollars).

The parametric estimated relationship in Tab. 1, represented in Fig. 1, shows that the significance of coefficients and explanatory power of model are high. The coefficient of $R^2$ adj. is also high and the model of CD as disruptive technology on established cassette technology explains more than 50% variance in the data (Tab. 1). In particular, results suggest that in the USA, CD technology has $B=2.1$ (i.e., $B>1$), suggesting a destruction of cassette technology with a *high relative rate of change* (period 1984-2008).

Tab. 2 and Fig. 2 show results of the disruptive streaming technologies on a period running from 2004 to 2018 =14 years. Streaming technology in this period is still in the phase of development with $B=-1.28$. In particular, the study shows that U. S. recorded music revenues of streaming technologies have overtaken CD technology in 2015 with $2,400 millions vs. $1,400 millions. This short period (i.e., 2015-2018, with 2018 last year under study here) may explain the relative rate of change of streaming technology vs. CD technologies in markets (that is $B<1$). Instead, the long period under study for CD technology−started in 1991 with recorded music revenues of $7,800 million of CD *vs.* $5,400 millions of cassette−explains the relative rate of substitution of CD as disruptive technology on cassette as established technology (i.e., $B>1$). In the context of streaming technology, Lee et al. (2016) find that online streaming services positively impact music record sales. Naveed et al. (2017) point out that the co-evolutionary pathways of increasing popularity of streaming services and resurgence of live music are sustaining music industry.

21 | P a g eCoccia M. (2020) *Cyclical phenomena in technological change*

*CocciaLab Working Paper 2020 – No. 54/2020*

In fact, the study by Naveed et al. (2017, original emphasis) on US music industry revealed that:

> (i) the co-evolution between streaming and live music industries has functioned well over the last few years, (ii) the live music industry has incorporated a self-propagating function by assimilating innovations previously initiated by digital music, (iii) given the above coevolution, the recent resurging trend in the music industry can be sustained, (iv) the advancement of digital innovations such as artificial intelligence, machine learning, fintech, virtual reality, big data, and social media by enabling such coevolution have transformed the live music industry into a 'live-concert streaming music industry'

## 1.3   Technological cycle of radical innovations

The second goal of this study stated in the introduction is to analyze technological cycle of disruptive innovations in industrial competition. The analysis of recorded music industry shows the evolution of different new technologies with different periods of beginning-ending of revenue in markets, as follows (cf., Tab. 3):

*Vinyl technology* (1930s-residual trend)⇒*8-track technology* (1965-1982)⇒*cassette technology* (1964-2008)⇒*CD technology* (1983-2018)⇒*Download technology* (2004-in progress) ⇒*streaming technology* (2005-in progress)

**Table 3** – Disruption period of established technology by new disruptive innovation in recorded music industry of the U.S. market

| Established Technology in market of recorded music | Year of the introduction of established technology | New disruptive innovation in market of recorded music | Year in which new technology destroys more than 50% of the revenue of established technology | % of recorded music revenues of established technology | Peak of revenues by established technology | Ending of revenues by established technology | Disruption Period (DP) in years of established technology via new disruptive innovation |
|---|---|---|---|---|---|---|---|
| Established | | Disruptive innovation | | | M | Z | $DP_i = Z_i - M_i$ |
| Vinyl | 1930 | 8-Track | n.a. | n.a. | 1979 | | |
| 8-Track | 1965 | Cassette | 1980 | 42.80 in 1980 | 1978 | 1982 | 4 |
| Cassette | 1964 | CD | 1991 | 41.00 in 1991 | 1990 | 2008 | 18 |
| CD | 1983 | Download | 2012 | 45.20 in 2012 | 2000 | 2018 | 18 |
| CD | 1983 | Download+ Streaming | 2011 | 46.60 in 2011 | | | |
| Download | 2004 | Streaming | 2015 | 49.98 in 2015 | | | |
| | | | Average values | 45.12% | | Average values | 13 years |
| | | | Standard Deviation (SD) | 3.47% | | SD | 8.8 years |

*Note*: elaboration on data by RIAA (2019, 2020); years are based on data concerning values not adjusted for inflation. Disruption Period of established technology $i$ is $DP_i = Z_i - M_i$ ($Z_i$ is the year of the ending of revenues of technology $i$ − $M_i$ is the year of the peak of revenues in technology $i$); n.a.=not available data.

Technology analysis of disruptive technologies suggests the following theoretical and empirical properties of technological change in the market under study here:

o *The property of average disruption period* states that a new technology destroys the established technology, overtaking total revenue in markets, in an average period of 13 years (SD=8.8years). *Remarks.* Results in Table 3 support this property showing the average duration of disruption period of established technologies in recorded music market.

The analysis of data by RIAA (2019, 2020) also shows different *technological cycles* driven by new disruptive innovations in U.S. recorded music market.

The *first technological cycle* is due to Vinyl records. The introduction of vinyl is in 1930 and the peak of revenues by vinyl single in U.S. recorded music market is in 1979 with €353,6M. In that year the sales volume for vinyl single format equaled 212,0M units, representing 31% of total sales volumes of 682.8M units for all formats that year. The introduction of cassette, CD and other technologies, as will be explained later, has almost destroyed this technology, such that in 2019 revenues by vinyl single are a mere €6,8M, representing a small niche within U.S. recorded music market. In addition, the vinyl format in 2019 is made up 0.1% of the total sales volumes of 453.3M units for all formats (RIAA, 2020). However, some scholars argue the diffusion of retro-technologies, such as vinyl record that is making a comeback, for rethinking the potential exploitation of their value and supporting value-creating strategies of physical retails in the digital age (cf., Saportg et al., 2016; Hracs and Jansson, 2017).

The *second technological cycle* is due to 8-track tape having the peak, measured with U.S. recorded music revenues, in 1978 (RIAA, 2019). However, in 1964 the cassette technology is also introduced in U.S. recorded market. This new technology has destroyed 8-track tape in 1982 with a disruption period of 4 years, given by difference between year of the ending of revenues of 8-track tape and year of its peak of revenues (i.e., 1982 −1978=4 years; cf., Tab. 3). The overall length of technological cycle of 8-track tape is 17 years (from 1965 to 1982; cf., Tab. 4).

The *third technological cycle* is due to cassette technology that started in U.S. recorded music market in 1964 and achieved the peak of revenue in 1990. However, cassette technology is destroyed in 2005 by (*new*) disruptive CD technology. The overall length of technological cycle of cassette technology is about 41 years, from the year of the starting of revenues to the year of the ending of revenues (cf., Tab. 4). Some scholars reveal the re-emergence of near-obsolete technologies, supporting a culture of vintage-technologies, such as vinyl records and cassettes (cf., Saportg et al., 2016; Eley, 2016). However, data show that technological cycles of these old-technologies are finishing, except some small niches of market associated with customers attached to vintage technologies (cf., Figures in Appendix).

The *fourth technological cycle* is given by CD technology that achieved the peak in 2001, after 18 years from its introduction in 1983. In 2018 this technology is almost destroyed by new technologies of download and video streaming. In fact, CD technology in 2018 has a mere $698.4 million of revenue on a total of $9,846 million in U.S. recorded music market. Moreover, the sales volume for CD format is made up 9.8% of the total volume of 532.3M units for all formats in 2018 (CD format had 91.1% of the total sales volumes of 968.5M units for all formats in 2001). The overall length of technological cycle of CD technology is about 35 years, whereas the disruption period is about 17 years (cf., Tabb. 3-4).

The *on-going technological cycles* in recorded music market are due to download and streaming technology introduced in the mid-2000s. However, download mode has had the peak in 2012, after 8 years from its introduction in 2004, and now it has a phase of decline because of streaming technology that is growing over time, driven by many technical advances, such as growing video-sharing websites (cf., Tab. 4).

Table 4 – Technological cycles in the U.S. recorded music industry

| Technological cycles in U.S. recorded music market | Up wave of cycle | | Down wave of cycle | Duration of technological cycle in years | | | | |
|---|---|---|---|---|---|---|---|---|
| | A begin of revenues of technology | **M** peak of revenues of technology | Z end of revenues of technology | AM length up wave years =M-A | MZ length down wave years [1] =Z-M | AZ length cycle years =Z-A | (AM / AZ) % | (MZ / AZ) % |
| 1  Vinyl technology | 1930 | 1979 | 2019 | 49 | 40 | 89 | 55.06 | 44.94 |
| 2  8-track tape technology | 1965 | 1978 | 1982 | 13 | 4 | 17 | 76.47 | 23.53 |
| 3  Cassette technology | 1964 | 1990 | 2005 | 26 | 15 | 41 | 63.41 | 36.59 |
| 4  CD technology | 1983 | 2001 | 2019 | 18 | 18 | 36 | 50.00 | 50.00 |
| 5  Download technology | 2004 | 2012 | * | 8 | - | - | - | - |
| 6  Streaming technology | 2005 | * | * | - | - | - | - | - |
| *Arithmetic mean, years* | | | | 22.80 | 19.25 | 45.75 | 61.24% | 38.76% |
| *Standard Deviation (SD), years* | | | | 16.08 | 15.09 | 30.63 | | |

Note. * technology in progress; elaboration on data from RIAA (2019); (1) Disruption period of established technology $i$ is $MZ_i$ = year of the ending of revenues of technology$_i$ ($Z_i$) − year of the peak of revenues of technology$_i$ ($M_i$); length of technological cycle of technology $i$ is $AZ_i$ = year of the ending of revenues of technology$_i$ ($Z_i$) − year of the starting of revenues of technology$_i$ ($A_i$).

These empirical results in Tab. 4 suggest another main property for disruptive innovations.

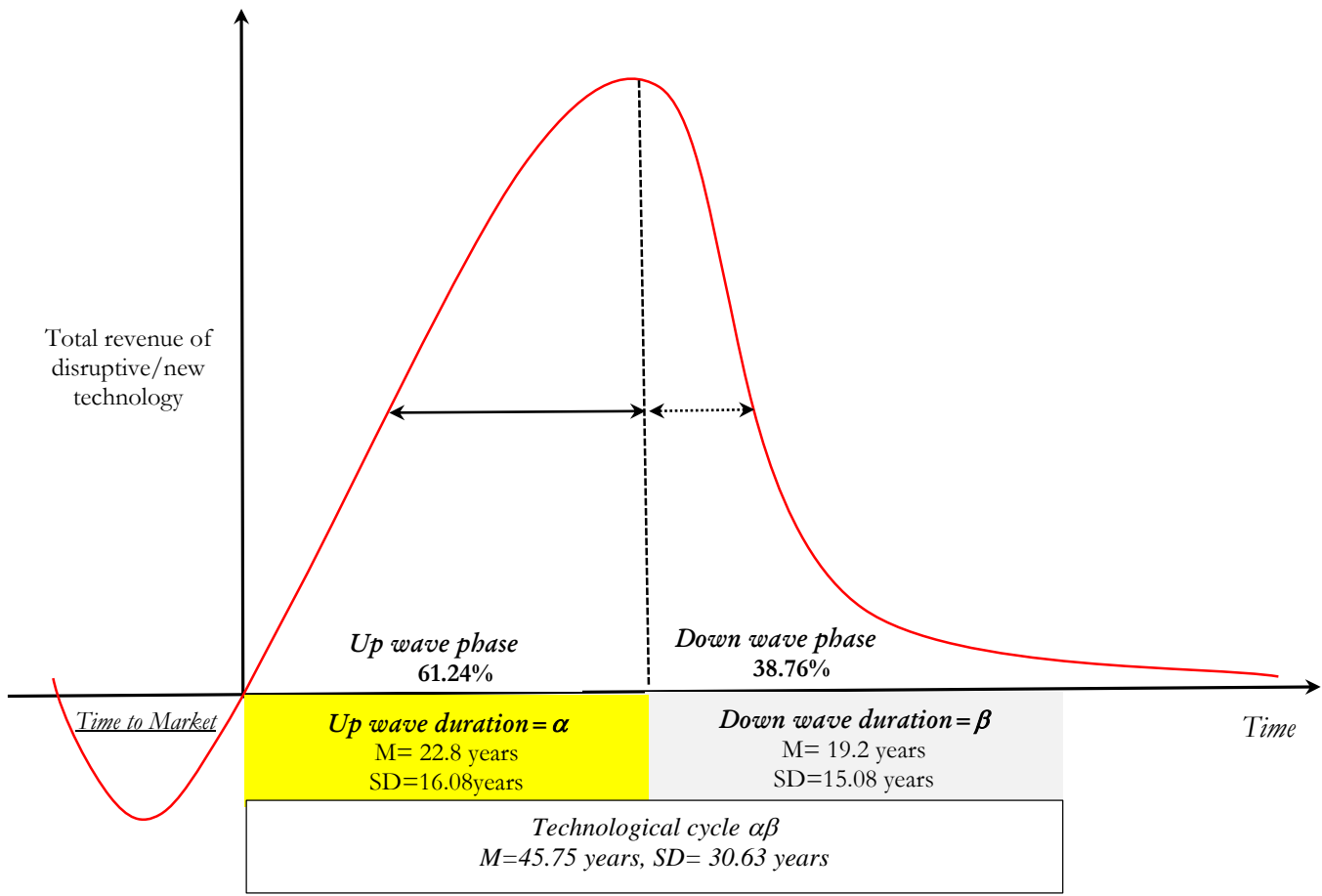

**Figure 3.** *Negative asymmetry of technological cycle of radical/disruptive innovations having up wave phase longer than down wave phase.*
*Note*: *M*=arithmetic mean based on historical data on the evolution of example technologies in recorded music industry of U.S. market; *SD* = Standard Deviation

o The property of the *asymmetry of technological cycle* states that technological cycle of disruptive innovations has up wave phase longer than down wave phase: *Negative asymmetry of technological cycle* (Figure 3).

*Remarks*. The analysis of technological cycles (vinyl, 8-track tape, cassette and CD technology) shows that up wave phase has an average duration of about 23 years (SD=16.8y), whereas down wave phase has an average duration of roughly 19 years (SD=15.1y). Average duration of technological cycle in recorded music market is about 46 years (Tab. 4). In particular, results suggest that technological cycles have an average duration of up wave phase equal to 61.24% of overall wavelength, whereas the average duration of down wave phase is shorter (it is about 38.76% of overall wavelength). Coccia (2010) showed that economic long waves have asymmetric paths with longer periods of *up wave* than *down wave* phase over time. A similar result seems to be also present in evolutionary cycle of research fields based on scientific production over time (cf., Coccia, 2020a). In general, this finding reveals an analogy between cycles of long waves, research fields and technologies that seem to have general socioeconomic factors supporting a longer phase of up wave over time (Coccia, 2010, 2020a, 2020b).

Hence, disruptive technologies have technical performance higher than established technologies, leading to a dominance in markets (cf., Christensen, 1997). This technological behavior of disruptive technology can be due to ambidexterity learning processes, given by:

− "*learning via diffusion*" (Sahal, 1981, p. 114, Italics added) in which the increased adoption of a technology supports the path for improvement in its technical characteristics (i.e., technological advances).

− "*diffusion by learning*" that improvement in the technical characteristics of a technology enhances the scope for its adoption over the course of time (Sahal, 1981, p. 114, Italics added).

**DISCUSSION AND IMPLICATIONS FOR MANAGEMENT OF TECHNOLOGY**

The competition between technologies is driven by a process of disruptive creation that generates technological and economic change over time (Calvano, 2006). Christensen (1997) argues that "disruptive technologies" offers a novel mix of attributes compared to established technology (cf., Christensen, 2006; Adner and Zemsky, 2005; Tellis, 2006). Adner and Zemsky (2005) show that disruption occurs when new-technology firms pursue a high-volume and low-price strategy that allow to break into the primary segment. However, the lower the costs of established-technology firms, the lower the threat of disruption. Adner and Zemsky (2005, p. 231) point out that: "On the one hand, the lowest-cost firm has the highest margins among new-technology firms, which favors output expansion and hence disruption. On the other hand, the lowest-cost firm has the highest market share in secondary segment and hence the most to lose from the fall in price that comes with disruption". Radnejad and Vredenburg (2019) suggest that to prevent failure in the process of developing and commercializing a potentially disruptive process innovation, new entrants in a process- oriented industry need to acquire three antecedents and capabilities given by: 1) Developing disruptive technological innovation requires alliance capability through all stages of development; 2) Effective intra-firm collaboration capability is an antecedent in developing a disruptive process innovation and 3) Participatory management is preferred leadership style in developing a radical technological process innovation in a process-oriented industry. Moreover, the development of a disruptive innovation requires a significant amount of capital that is unlikely to be available to new entrants or smaller companies. The R&D of disruptive innovation can be supported by a collaboration strategy that protects the rights of small companies in industries while making it worthwhile for incumbents to participate and invest. In this context, new entrant needs to develop alliance capability, inter-organizational capability and participatory management style as well as to secure investment from governments that understand the timeline of developing disruptive innovations. Adner and Zemsky (2005) also argue that social welfare can increase because prices for products fall with disruption and that concentration tends to increase with disruption because the effect of cost asymmetries on market share is amplified by the increased number of competitors. In this context, the results of proposed theoretical and empirical analyses are that:

1. The growth of disruptive technology is generally an allometric process of *a disproportionate growth* of (new) disruptive technology in relation to the established technology
2. The technological cycle of disruptive innovations has up wave phase longer than down wave phase (*asymmetry of technological cycle*)
3. The ambidexterity learning processes, based on *learning via diffusion* and *diffusion by learning,* are a driving force underlying the development and adoption of disruptive technology in turbulent (complex and fast changing) markets
4. Finally, technologies can have a higher capacity of disruption if they are inter-related with other technologies as host or parasitic technologies, such as streaming technologies (cf., Coccia, 2018a; Coccia, 2019, 2019a, 2019b; Coccia and Watts, 2020).

To our knowledge, this study is the first attempt to develop a specific allometric model that measures the growth rate of (new) *disruptive technologies* in markets during a competition with existing technologies. This study can provide best practices of strategic management to predict the directions and pathways of disruptive technologies based on *relative growth rate* **B** of proposed model (cf., Krotov, 2019). In fact, coefficient **B** of technological disruption model can support innovation strategy of firms on critical decisions of when to invest in Research & Development (R&D) of new disruptive technologies, abandon the old technology or pursue an intermediate level of R&D investment between old and new technology for sustaining and safeguarding competitive advantage in turbulent markets. In addition, Christensen (1997) observes that established firms face an *innovator's dilemma* concerning internal resource-allocation processes that lead them to systematically underinvest in disruptive technologies (cf. also, Christensen and Bower, 1996). Kapoor and Klueter (2015) also show that incumbents tend to not invest in disruptive technological regimes and maintain a competence-enhancing approach. However, research alliances and acquisitions are strategies that may help incumbents to overcome this *inertia* both in the initial stage of research and in the later stage of development of disruptive technologies (cf., Radnejad and Harrie, 2019). A main example is represented by the strategic alliance between AstraZeneca (incumbent) and Amgen (a leader in biotechnology) to co-develop and commercialize five monoclonal antibodies from Amgen's clinical portfolio. The alliance with Amgen is designed to share risk and leverage each partner's functional and geographic strengths. This collaboration improves the expertise of AstraZeneca in respiratory and gastrointestinal diseases, whereas Amgen improves commercial experience in rheumatologic and dermatologic diseases (Coccia, 2014, p. 742).

In general, disruptive technologies compete with established technologies to achieve the dominance in markets, generating industrial and corporate change. For scholars, disruptive technologies highlight the question of the boundaries of technology competition and how those boundaries change over time (Adner, 2002). For managers, disruptive technologies highlight the danger posed to incumbent firms from too quickly dismissing new technologies as inferior and therefore irrelevant to their market positions. Overall, then, the study here has tried to explain, whenever possible, new characteristics and properties of the behavior of disruptive technologies that can expand disruptive innovation theory. However, we know, *de facto*, that other things are often not equal over time and space in the domain of technology. The study here may encourage further theoretical exploration in the *terra incognita* of the competition between technologies that generates a disruptive creation for technological and economic change in society. Future efforts in this research field will be also directed to provide further empirical evidence, also considering dependency-network framework between technologies to better explain the behavior of disruptive technologies in different markets (cf., Mazzolini et al., 2018; Iacopini et al., 2018). To conclude, identifying a generalizable theory of disruptive technologies for prediction of rapid change in industrial competition is a non-trivial exercise. In fact, Wright (1997, p. 1562) properly claims that: "In the world of technological change, bounded rationality is the rule."

**Appendix:** Technological cycles in the U.S. Recorded Music industry 1973-2019. *Source* RIAA (2020)

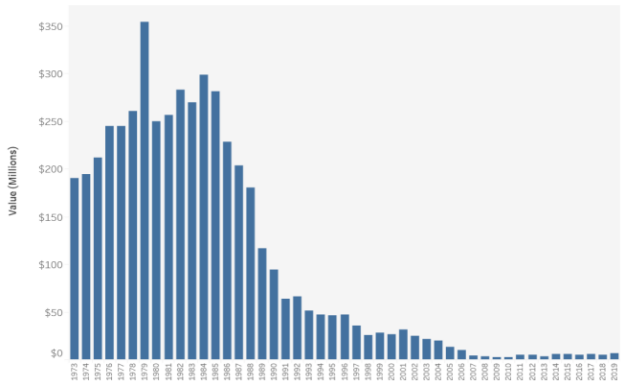
**Vinyl technology 1973-2019**

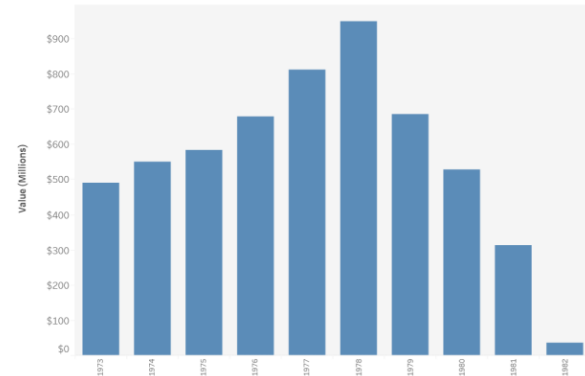
**8-track technology 1973-1982**

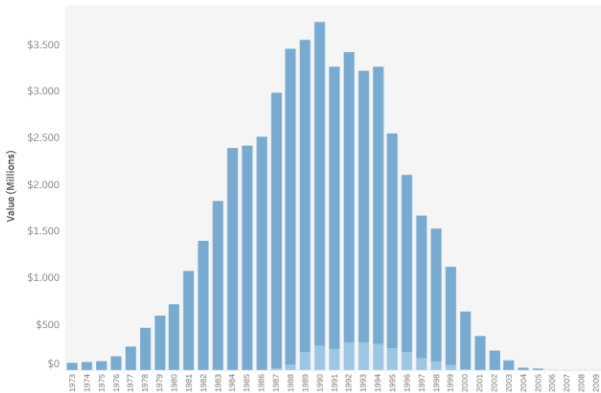
**Cassette technology 1973-2005**

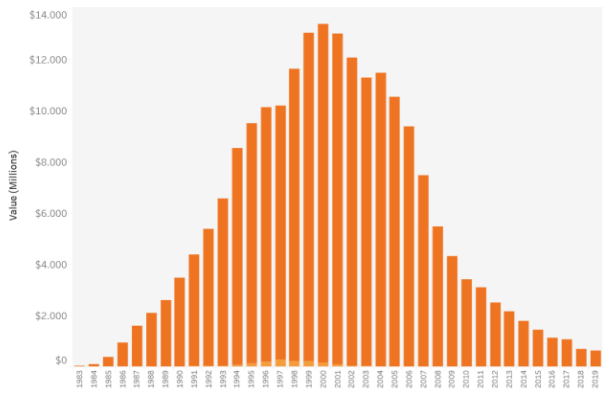
**CD technology 1983-2019**

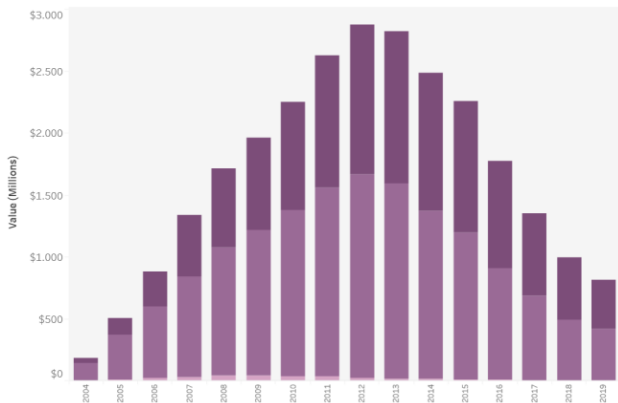
**Download technology 2004-2019**

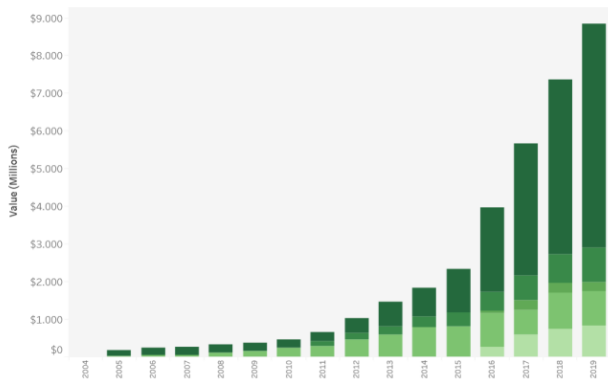
**Streaming Technologies 2004-2019**

33 | P a g eCoccia M. (2020) *Cyclical phenomena in technological change*

*CocciaLab Working Paper 2020 – No. 54/2020*

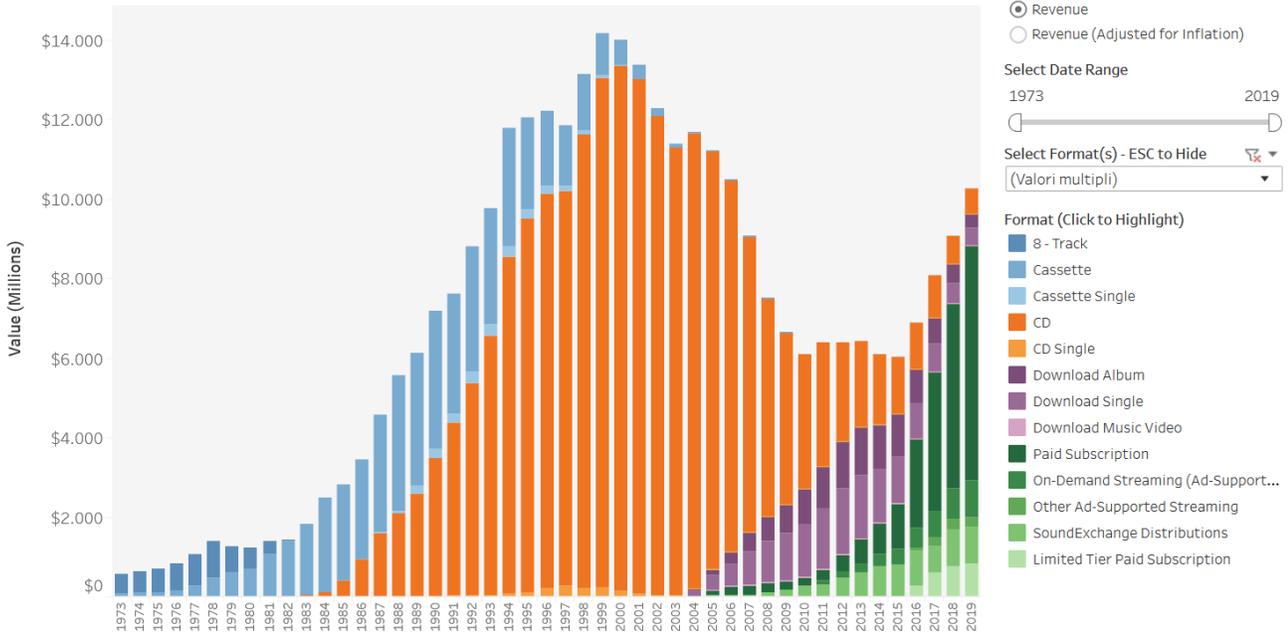

**Figure 1A.** Technological waves in recorded music industries from 1973 to 2019.